\documentclass[aps,prd,showpacs,preprintnumbers,nofootinbib,floatfix,floats,groupedaddress,twocolumn]{revtex4}

\usepackage{bm}
\usepackage{latexsym}
\usepackage{dcolumn}
\usepackage{amsmath,amsfonts,amssymb}
\usepackage{graphicx,epsfig}

\def \l{\left}
\def \r{\right}
\def \DM{\mathrm{d}}
\def \LRL {Laplace-Runge-Lenz }

\newcommand{\nv}{\hat{\mathbf{n}}}

\newcommand{\rv}{\mathbf{r}}

\reversemarginpar

\begin{document}
\title{Duality of force laws and Conformal transformations}
%
\author{Dawood Kothawala}
\email{dawood.ak@gmail.com, dawood@iucaa.ernet.in}

\affiliation{Department of Mathematics and Statistics, University of New Brunswick, Fredericton, NB, Canada E3B 5A3}

\date{\today}
\begin{abstract}
As was first noted by Isaac Newton, the two most famous ellipses of classical mechanics, arising out of the force 
laws $F \propto r$ and $F \propto 1/r^2$, can be mapped onto each other by changing the 
location of center-of-force (CoF). What is perhaps less well known is that this mapping can also be achieved by 
the complex transformation, $z \rightarrow z^2$. We give a simple derivation of this result (and its generalization) 
by writing the Gaussian curvature in its ``covariant" form, and then changing the \textit{metric} by a conformal 
transformation which ``mimics" this mapping of the curves. The final result also yields a relationship between Newton's constant $G$, mass $M$ of the central attracting body in Newton's law, the energy $E$ of the Hooke's law orbit, and the angular momenta of the two orbits.

We also indicate how the conserved Laplace-Runge-Lenz 
vector for the $1/r^2$ force law transforms under this transformation, and compare it with the corresponding quantities for the linear force law. Our main aim is to present this duality in a geometric fashion, by introducing elementary notions from differential geometry. 
\end{abstract}

\maketitle
\vskip 0.5 in
\noindent
\maketitle
\section{Introduction} \label{sec:intro}
In the \textit{Principia}, Newton discusses the following problem \cite{newton, chandra}: Given a trajectory in a 
particular force field, how must the CoF and the force law be changed so that one obtains, for the new situation, 
the same orbit as before. Newton, of course, attacked the problem purely geometrically and showed that doing so takes 
one from a linear force law [\textit{Hooke's} law] with CoF at the origin, to an inverse-square force law with CoF at 
one of the foci of the ellipse. This problem can be approached in an alternate way by fixing the CoF at the origin and 
changing the orbit such that the new orbit is a valid trajectory under a modified force law; this 
naturally allows one to use the powerful techniques of complex analysis. This has been discussed in detail by Needham in his wonderful book \cite{needham-1}, and also in \cite{needham-2}. In this pedagogical note, we prove the result in a somewhat different and amusing manner by 
mapping the problem to an equivalent one in which the curve remains fixed, but the underlying metric undergoes a 
conformal transformation thereby changing the Gaussian curvature of the curve. Since the force depends on $\kappa$, this immediately leads to the desired result as well as to its generalization.

{\it Notation}: We shall set the mass $m$ to unity for ease of notation. Our convention for tangent and normal vectors can be simply stated by giving these for a circular trajectory: in our convention, a circular trajectory has tangent vector pointing towards increasing $\theta$ (i.e., moving counter-clockwise) and the normal is {\it outward} unit radial vector. While intermediate results might depend on convention, the final physical results are, of course, independent of it.
\section{The Gaussian curvature of a curve} \label{sec:gauss-curv}
Before we discuss in detail our definition of curvature of a curve, it is worth recalling the precise relationship between the force generating a trajectory and the curvature of the trajectory.

Let us begin with Newton's law: $\bm F = \DM^2 \bm r/\DM \tau^2$, where $\tau$ is standard time coordinate. Now, if $s$ denotes the arc-length parameter along the trajectory, then we can write $\DM \bm r = \bm t \DM s$, where $\bm t$ is the unit tangent vector to the trajectory. Defining the speed of the trajectory as $v=\DM s/\DM \tau$, we can write $\bm F = v^2 \DM \bm t / \DM s + \bm t (\DM v /\DM \tau)$. This relation forms the basis for analysis of Newton's law by breaking the force into components parallel and normal to the trajectory. Note that $\bm t \cdot \DM \bm t / \DM s=0$ since $t^2=1$, which allows us to define a unit normal vector $\bm n$ by $\DM \bm t/\DM s=-\kappa \bm n$, where $\kappa$ is called the Gaussian curvature \cite{comment-1}. We therefore have:
\begin{eqnarray}
F &=& \bm n F_{_N} + \bm t F_{_T} \label{eq:force-fn-ft}
\\
F_{_N} &=& \bm F \cdot \bm n = - \kappa v^2 \; ; F_{_T} = \bm F \cdot \bm t =  \frac{\DM v}{\DM \tau}
\end{eqnarray}
For central force laws of the form $\bm F = - F \hat{\bm r}$, we obtain (refer Figure \ref{fig:set1}): $F \cos \gamma = \kappa v^2$. We shall use this relation later in Section (\ref{sec:dual-force}) [see Eq.\;(\ref{eq:Fnormal})].

Let us now return to the discussion of the curvature itself. A standard result from high school analytical geometry gives the expression for $\kappa$ of a curve, $y=f(x)$, as \cite{thomas-cal}
\begin{eqnarray}
\kappa = - \frac{f''}{(1+f'^2)^{3/2}} \label{eq:kappa}
\end{eqnarray}
where the choice of overall sign depends on convention; we will explain our choice below. At a slightly advanced level, one recognizes $\kappa$ as related in some way to the ``extrinsic curvature"  \cite{mtw-Kab} of the curve [since a curve is one-dimensional, it has no intrinsic curvature of its own]. Since the extrinsic curvature depends on embedding of the curve in the Euclidean 
plane, it will change not only when we deform the curve, but also when we change the background geometry. This is 
essentially the trick we will use to relate $\kappa$'s associated with two different curves: we simple change the 
background geometry in an appropriate manner, \textit{while keeping the curve fixed}. The advantage of doing it this 
way is that, if one has available a covariant expression for $\kappa$, one can easily evaluate the change in 
$\kappa$ purely due to conformal change of the metric.

Consider the expression for the divergence, $\mbox{div}\,[\nv]$, of the vector $\nv$ which is the unit normal to the curve. For $y=f(x)$, the unit normal is given by ${\hat n}_k=||n||^{-1} \partial_k [y-f(x)]=(-f',1)/||n||$, where $||n|| = \sqrt{1+f'^2}$. Note that the normal vector is most naturally defined with lower indices, that is, as a co-variant vector (technically, a \textit{one-form}). This fact, although not relevant in Cartesian coordinates, is crucial in arbitrary coordinates in which the metric is non-trivial. The divergence of $\nv$ is then seen to give $\kappa$ in Eq.~(\ref{eq:kappa}) with the minus sign; that is,
\begin{eqnarray}
\kappa = \mbox{div} [ \nv ]
\label{eq:kappa-def-N}
\end{eqnarray} 
If we evaluate the expression for a circle of radius $a$, the outward unit normal is $\hat{\rv}$, and we obtain $\kappa=1/a$. For an arbitrary trajectory in a given force law, we require the curvature associated with the \textit{osculating circle} at a given point. Because it is natural to choose the outward normal which gives the standard result for the circle, we choose the minus sign in Eq.~(\ref{eq:kappa}).

\textit{Aside:} Although this definition of $\kappa$ is not usually encountered in standard texts, those working with differential geometry and general relativity would recognize it as the trace of the extrinsic curvature tensor, $K_{ab}$, and a direct connection is established in Appendix~\ref{app:kappa-and-ext-curv}. Because the expression (\ref{eq:kappa-def-N}) is in a tensorial form, we can directly use it for any metric. However, there is a subtlety that is worth mentioning for readers with a background in advanced geometry. The conformal mappings we are considering are \textit{diffeomorphisms} of the complex plane,\cite{comment-2} and we expect to exploit the obvious advantage of \textit{generally covariant} expressions to make meaningful statements about important characteristics of a manifold or a submanifold. However, it is well known that any 2-dimensional space can be written in a conformally flat form, and hence our mappings cannot generate all possible equivalence classes of metrics related to the Euclidean metric by conformal transformations. A $2$-sphere is an instructive example (we know that a $2$-sphere is not equivalent to a flat space under diffeomorphisms). Because a $2$-sphere is genuinely curved, any statement about the curvature of curves on a $2$-sphere depends on whether we look upon that curve as being embedded in the $2$-sphere, or in flat $3$-dimensional space. We invite the reader to read Appendix~\ref{app:embedding-of-curves} for more information. 

\section{Conformal transformation of $\kappa$} \label{sec:conf-transf}
We shall now derive how the Gaussian curvature changes under a conformal mapping of curves. That is, we want to find  the Gaussian curvature $\bar \kappa$ of the curve $\mathcal{\bar C}$ which is obtained from a curve ${\mathcal C}$ (with curvature $\kappa$) by a conformal mapping, $\Phi: \mathcal{C} \rightarrow \mathcal{\bar C}$. Hence, a point $z$ on ${\mathcal C}$ is mapped to a point $w=\Phi(z)$ on $\mathcal{\bar C}$. First, consider two infinitesimally separated points $P$ and $Q$ on the curve $\mathcal{C}$; the distance between them is $\DM s^2 = \DM x^2 + \DM y^2 = \DM z \DM z^{*}$, where the second equality expresses the interval in terms of the complex variable, $z=x+iy=r \exp{i \theta}$. The map $\Phi$ takes a point $z$ to $w=\Phi(z)$. Therefore, $P$ and $Q$ are mapped respectively to $\bar P$ and $\bar Q$, and the line interval between these is given by, $\DM {\bar s}^2 = |\Phi'(z)|^2 \DM z \DM z^{*}$. That is, the line intervals are related by a conformal transformation: $\DM {\bar s}^2 = \Omega(x,y)^2 \DM s^2$, with $\Omega = |\Phi'(z)|$. 

We shall now give two derivations relating $\bar \kappa$ to $\kappa$. The first one is short and elegant, while the second derivation, although lengthy, is more along conventional lines; these derivations, specifically the first one, are not easily found in standard texts.

{\bf Method 1}: Consider two line elements, related by a conformal transformation: $\DM \bar{s}^2 = \Omega(x,y)^2 \; \DM s^2$, where $\DM s^2 = \DM x^2 + \DM y^2$. Note that the metric tensor ${\bar g}_{ab}$, defined by $\DM {\bar s}^2={\bar g}_{ab} \DM x^a \DM x^b$, is ${\bar g}_{a b} = \Omega^2 \delta_{a b}$ and its inverse ${\bar g}^{a b} = \Omega^{-2} \delta_{a b}$.

The equation $y=f(x)$ describes the curve $\mathcal{C}$ with line element $\DM s^2$, and also the curve $\mathcal{\bar C}$ with line element $\DM \bar{s}^2$. That is, instead of actually changing the curve, we keep the curve fixed and change the background metric from $\delta_{ab}$ to ${\bar g}_{ab}$. We shall now use the definition of $\kappa$ in terms of divergence of the normal vector div[$\hat{\bm n}$], motivated in the previous section, to relate $\bar \kappa$ with $\kappa$.

For a curve described by $y=f(x)$, the unit normal vector, call it $\bm \hat{\bm \ell}$, is given by:
\begin{eqnarray}
\hat{\ell}_{k} = \frac{\Omega (-f',1)}{\sqrt{1+f'^2}}  \; ; \;\; \hat{\ell}^{k} = {\bar g}^{ki} \hat{\ell}_{i} = \frac{(-f',1)}{\Omega \sqrt{1+f'^2}} 
\equiv \frac{1}{\Omega} n^{k}
\label{eq:vec-rel}
\end{eqnarray}
Note that ${\bar g}_{ab} \hat{\ell}^{a} \hat{\ell}^{b} = 1$. (The $\equiv$ above is only exhibited for further {\it computational convenience}). Then, by definition:
\begin{eqnarray}
\bar{\kappa} = \mathrm{div}[\hat{\bm \ell}] &=& \frac{1}{\sqrt{\bar g}} \partial_{k} \left( \sqrt{\bar g} \; \hat{\ell}^{k}  \right) \\
&=& \frac{1}{\Omega^2} \partial_{k} \left( \Omega^2 \; \hat{\ell}^{k}  \right)
\end{eqnarray}
where we have used the well known expression for the divergence in terms of metric tensor. Here, $\bar g$ is the determinant of ${\bar g}_{a b}$. Clearly, ${\bar g} = \Omega^4$, which we have substituted in the second
line. Using the second of Eqs.~(\ref{eq:vec-rel}), we immediately obtain the desired relation:
\begin{eqnarray}
\bar{\kappa} = \Omega^{-1} \kappa + \Omega^{-2} \; \hat{n}^{k} \partial_{k} \Omega
\label{eq:kappa-rel}
\end{eqnarray}
This is an extremely useful relation; on the LHS, we have the Gaussian curvature of a curve $\mathcal{\bar C}$ which 
is obtained from a curve ${\mathcal C}$ by a conformal mapping, while on the RHS, all the quantities refer to the 
original curve $\mathcal C$, which, in our construction, is a curve in a simple, Cartesian system of coordinates. 

The main aim of this note was to exhibit the simple derivation of Eq.~(\ref{eq:kappa-rel}) using very elementary 
differential geometry and conformal transformation [we are not aware of any standard text containing this derivation]. 
To reiterate, we have just mapped the problem of finding the Gaussian curvature of a deformed curve in terms of that 
of the original curve, to another problem in which we consider the same curve but in a different metric. Essentially, 
our trick is based on the fact that all the information about the shape of a curve is encoded in the infinitesimal distance between two points on the curve, and these infinitesimal distances for the two curves are related by a conformal factor. 

However, since the covariant expression for divergence of a vector may not be familiar to some readers, we give below a brute force derivation of this same result, which is lengthy but otherwise straightforward.

{\bf Method 2}: We now deal with the conformal mapping directly, by mapping each point $z$ on curve $\cal C$ to $w=\Phi(z)$ on $\cal{\bar C}$. Note that $\cal C$ and $\cal{\bar C}$ are now two different curves in the same Euclidean plane. Then, the unit tangent and normal vectors $\bm t$ and $\bm n$ at point $P$ on $\cal C$ go to $\bm T$ and $\bm N$ at $\Phi(P)$ on $\cal{\bar C}$. Using the conditions: $T^2=1=N^2, \bm T \cdot \bm N=0$, we can write the relation between these vectors as [$\epsilon=\pm 1$]:
\begin{eqnarray}
\bm T &=& \alpha \bm t + \epsilon \sqrt{1-\alpha^2} \bm n 
\\
\bm N &=& \alpha \bm n - \epsilon \sqrt{1-\alpha^2} \bm t
\end{eqnarray}
where $\alpha = \bm T \cdot \bm t = \cos \phi, \phi$ being the angle between $\bm T$ and $\bm t$ 
\cite{comment-3}.
It is then straightforward to show that [$\Omega=|\Phi'(z)|$]:
\begin{eqnarray}
\bar \kappa = - \bm N \cdot \frac{\DM \bm T}{\DM \bar s}  = \frac{1}{\Omega} \l[ \kappa - \epsilon \frac{\DM \phi}{\DM s} \r] 
\label{eq:kappa-rel-ref1}
\end{eqnarray} 
Hence, our problem reduces to the task of finding ${\DM \phi}/{\DM s}$, which turns out to be daunting. We shall use the machinery of complex analysis to simplify analysis as much as possible. We write $w=\Phi(z)=u(x,y)+i v(x,y)$ and use the Cauchy-Riemann equations (the map $\Phi$ is assumed to be analytic): $u_x=v_y, u_y=-v_x$, where $u_x \equiv \partial u / \partial x$ and so on. As a consequence, we also have $u_{xx}+u_{yy}=0=v_{xx}+v_{yy}$. Further, we shall represent the vectors $\bm t, \bm T$ in complex notation as: $t=\dot x + i \dot y$, $T=\Omega^{-1}(\dot u + i \dot v)$, where dot stands for $\DM/\DM s$ and we have used $\DM \bar s=\Omega \DM s$. Also, the condition $\bm t \cdot \bm n=0$ leads to $n=-i t=\dot y - i \dot x$ (the minus sign is in accord with our sign convention described in Sec.~I). We now obtain:
\begin{eqnarray}
T = \Omega^{-1} \l( \dot u + i \dot v \r) = \alpha \l( \dot x + i \dot y \r) + \epsilon \sqrt{1-\alpha^2} \l( \dot y - i \dot x \r)
\end{eqnarray}
Equating the real and imaginary parts give the relations: $\alpha=\Omega^{-1} u_x$ and $\epsilon \sqrt{1-\alpha^2}=\Omega^{-1} u_y$. These further lead to following useful identities: (i) $\dot \alpha = - \sqrt{1-\alpha^2} \dot \phi=-\epsilon \Omega^{-1} u_y \dot \phi$, where the first equality follows from definition of $\alpha$, and (ii) differentiating $\alpha=\Omega^{-1} u_x$, and using chain-rule, we obtain:
\begin{eqnarray}
\frac{\DM \alpha}{\DM s} = -\frac{\dot \Omega}{\Omega^2} u_x + \frac{1}{\Omega} \l( \dot x u_{xx} + \dot y u_{yx} \r)
\end{eqnarray}
Combined with  $\dot \alpha = -\epsilon \Omega^{-1} u_y \dot \phi$, this gives
\begin{eqnarray}
\epsilon \frac{\DM \phi}{\DM s} = \l( \frac{u_x}{u_y} \r) \frac{\DM \ln \Omega}{\DM s} - \frac{\dot x u_{xx} + \dot y u_{yx}}{u_y}
\label{eq:kappa-rel-ref2}
\end{eqnarray}
This relation can be simplified considerably and expressed directly in terms of derivatives of the mapping $\Phi(z)$, which will help us to recast Eq.\;(\ref{eq:kappa-rel-ref1}) in the form (\ref{eq:kappa-rel}). To proceed, we first note that $\partial_z=(1/2)(\partial_x-i \partial_y)$, which follows by writing $x=(z+z^*)/2, y=(z-z^*)/2i$, and treating $z$ and $z^*$ as independent. Operating on $\Phi(z)$, and using the Cauchy-Riemann relations, we obtain:
\begin{eqnarray}
\Phi'(z) &=& \partial_z \Phi(z) = u_x - i u_y
\\
\Phi''(z) &=& \partial_z^2 \Phi(z) = u_{xx} - i u_{yx}
\end{eqnarray}
from which it follows that: $t  \Phi''(z) = \l( \dot x u_{xx} + \dot y u_{yx} \r) + i \l( \dot y u_{xx} - \dot x u_{x y} \r)$. Therefore, ${\rm Re}[t \Phi'']$ is precisely what appears in the second term in the expression for $\dot \phi$. The first term is also easily simplified by noting that, $\DM \ln \Omega/\DM s = \bm t \cdot \nabla \ln \Omega = {\rm Re}[t (\nabla \ln \Omega)^{*}]$. Now, define $\Phi'=F$, so that $\Omega=|\Phi'(z)|=\sqrt{F(z) F^*(z^*)}$. Therefore, $\ln \Omega=(1/2) \l( \ln F + \ln F^* \r)$. Also, gradient of a real function is represented as: $\nabla \Omega=\partial_x \Omega + i \partial_y \Omega=2 \partial_{z^*} \Omega(z,z^*)$. Hence, $\nabla \ln \Omega=\partial_{z^*} \ln F^*(z^*)/\partial z^*$, so that we finally arrive at: $(\nabla \ln \Omega)^{*}=\partial_z \ln F(z)/\partial z=\Phi''/\Phi'$. Putting everything together, we finally obtain: $\DM \ln \Omega/\DM s = {\rm Re}[t \Phi''/\Phi']$. Therefore, Eq.\;(\ref{eq:kappa-rel-ref2}) simplifies to:
\begin{eqnarray}
\epsilon \frac{\DM \phi}{\DM s} &=& \frac{1}{u_y} \l( u_x {\rm Re}\l[\frac{t \Phi''}{\Phi'}\r] - {\rm Re}[t \Phi''] \r)
\\
&=& - \frac{1}{{\rm Im}[\Phi']} \l( {\rm Re}[\Phi'] {\rm Re}\l[\frac{t \Phi''}{\Phi'}\r] - {\rm Re}[t \Phi''] \r)
\end{eqnarray}
We now use the fact that, for any two complex numbers $p$ and $q$: ${\rm Re}[p q]={\rm Re}[p] {\rm Re}[q] - {\rm Im}[p] {\rm Im}[q]$. Therefore,
\begin{eqnarray}
\epsilon \frac{\DM \phi}{\DM s} = - {\rm Im}\l[\frac{t \Phi''}{\Phi'}\r]
\end{eqnarray}
Substituting in (\ref{eq:kappa-rel-ref1}), we obtain
\begin{eqnarray}
\bar \kappa = \frac{1}{\Omega} \l[ \kappa + {\rm Im}\l[\frac{t \Phi''}{\Phi'}\r] \r] 
\label{eq:needham-kappa}
\end{eqnarray}
It is now straightforward to relate this to Eq.\;(\ref{eq:kappa-rel}) by using $t=i n$, $\Phi''/\Phi'=(\nabla \ln \Omega)^*$, and noting that for any two complex numbers $p$ and $q$, ${\rm Re}[p q^*]$ gives the dot product between vectors $\bm p$ and $\bm q$ represented by these complex numbers:  ${\rm Re}[p q^*] = \bm p \cdot \bm q$. Hence, ${\rm Im}[{t \Phi''}/{\Phi'}] = {\rm Re}[n (\nabla \ln \Omega)^*] = \bm n \cdot \bm \nabla \ln \Omega$. Therefore, we obtain
\begin{eqnarray}
\bar \kappa = \Omega^{-1} \l[ \kappa +  \hat{n}^{k} \partial_{k} \ln \Omega \r] 
\end{eqnarray}
which is the same as Eq.\;(\ref{eq:kappa-rel}).

Before proceeding, we must mention that an alternative derivation of Eq.\;(\ref{eq:needham-kappa}) has been given by T. Needham \cite{needham-1, needham-2} using elegantly a combination of pictorial and geometric methods in complex analysis.

\section{Duality of forces} \label{sec:dual-force}
\begin{figure}[!htb]
\begin{center}
\resizebox{400pt}{300pt}{\includegraphics{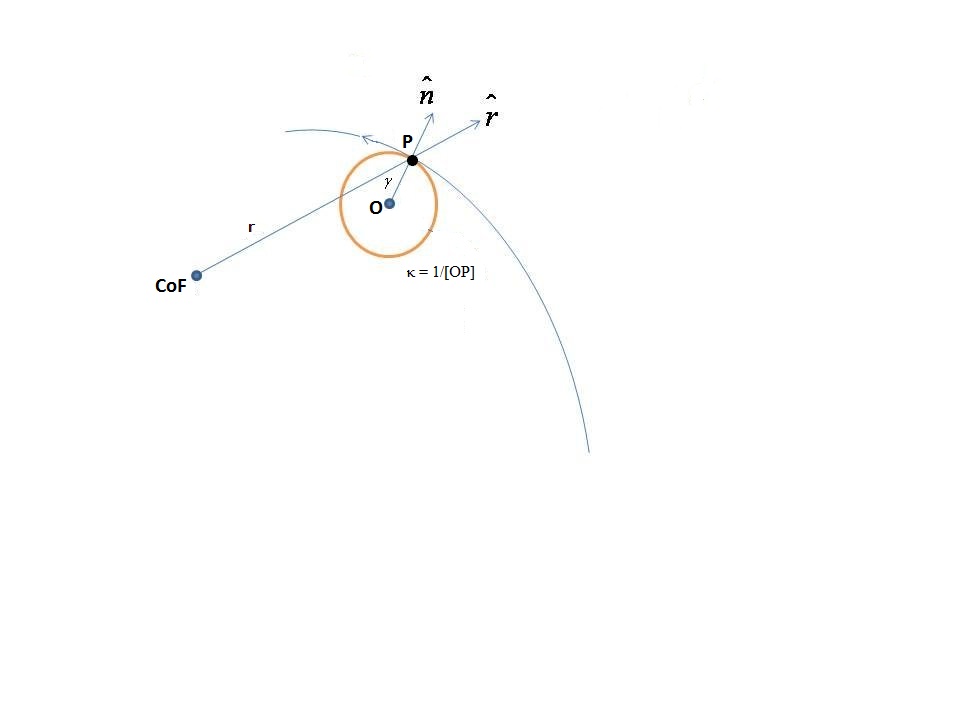}}
\vskip -140 pt
\vskip 50 pt
\caption{The figure shows relevant geometric quantities associated with a curve. For a point $P$ on the curve, $\hat{\bm n}$ is the unit normal to the curve (and also to the osculating circle) at $P$, $\hat{\bm r}$ is the unit vector directed from CoF to $P$; $r$ is the distance between CoF and $P$. Also, $\cos \gamma=\hat{\bm n} \cdot \hat{\bm r}$. By definition, radius of the osculating circle is $[OP]=1/\kappa$, $\kappa$ being the curvature at $P$.}
\label{fig:set1}
\end{center}
\end{figure}

We shall now outline the derivation of duality between force laws using results obtained above, along the lines of \cite{needham-1,needham-2}. We refer to Figure \ref{fig:set1} for the geometric quantities associated with the curve.

Let us recall a few elementary results from analytic geometry. We have,
\begin{eqnarray}
F \cos{\gamma} &=& \kappa v^2  \label{eq:Fnormal} \\
h &=& | \bm r \times \bm v | = r v \cos{\gamma}
\end{eqnarray}
which imply
\begin{eqnarray}
F = \frac{h^2 \kappa \sec^3{\gamma}}{r^2}
\end{eqnarray}
where $h$ is the (conserved) angular momentum. For definiteness, we have assumed the force to be directed towards the 
center. Our aim is to find the new force, $\bar F$, under which the deformed curve $\mathcal{\bar C}$ is a trajectory. 
So we need to find how all the terms on the RHS of the above equation transform. Since the mapping is conformal, it 
preserves angles, so that $\bar \gamma = \gamma$. Further, since $r=|z|$, its transformation is fixed by the mapping. 
Finally, the transformation rule for $\kappa$ is given by Eq.~(\ref{eq:kappa-rel}). Although Eq.~(\ref{eq:kappa-rel}) 
is general, we shall concentrate on specific mappings of the form, $\Phi(z)=z^m$. Therefore, we have, 
$\Omega = |\Phi'(z)| = m \; r^{m-1}$. In particular, we have, $\bm \nabla \Omega = m (m-1) r^{m-2} \bm \hat{\bm r}$, 
thereby giving $\bm \hat{\bm n} \cdot \bm \nabla \Omega = m (m-1) r^{m-2} \cos{\gamma}$ which trivially follows from Figure \ref{fig:set1}. Using this to simplify the right-hand side of Eq.~(\ref{eq:kappa-rel}), we obtain
\begin{eqnarray}
\bar \kappa &=& \frac{1}{m r^{m-1}} \l( \kappa + (m-1) \frac{\cos{\gamma}}{r} \r) \\
\nonumber \\
\bar F &=& \frac{{\bar h}^2 \bar \kappa \sec^3{\bar \gamma}}{{\bar r}^2} \\
&=& \l( \frac{\bar h}{h} \r)^2 \frac{1}{{\bar r}^2} \frac{2 (m-1)}{m} \l[ \frac{1}{2 (m-1)} r F + \frac{1}{2} v^2 \r] r^{2-m}
\nonumber \\
\end{eqnarray}

Now, suppose that $\bm F = - F \hat{\bm r} = - k r^A \hat{\bm r}$ ($k>0$). This force can be derived from a potential, 
$\phi = r F / (1+A)$. 
\begin{eqnarray}
\bar F = \l( \frac{\bar h}{h} \r)^2 \; {\bar r}^{(2-3m)/m} \; \frac{2 (m-1)}{m} \l[ \frac{1+A}{2 (m-1)} \phi  + \frac{1}{2} v^2 \r]
\nonumber \\
\end{eqnarray}
Now, \textit{if} $1+A = 2 (m-1)$, so that $m=(A+3)/2$, we obtain,
\begin{eqnarray}
\bar F &=& \l( \frac{\bar h}{h} \r)^2 \; {\bar r}^{\bar A} \; \frac{2 (m-1)}{m} \l[ \phi  + \frac{1}{2} v^2 \r] \\
&=& \l\{ E \l( \frac{\bar h}{h} \r)^2 \frac{2 (m-1)}{m} \r\} \; {\bar r}^{\bar A} \label{eq:force-law}
\\
&\propto& {\bar r}^{\bar A}
\end{eqnarray}
where $E$ is the (conserved) energy of the original orbit, and $\bar A$ is given by, $(A+3)(\bar A + 3)=4$. This 
completes the derivation. To summarize, the force laws $F$ and $\bar F$ are dual under a conformal mapping $z^m$ 
if $m=(A+3)/2$; the power law indices are then related as $(A+3)(\bar A + 3)=4$. The case, $A=1, m=2$ maps Hooke's 
law to Newton's inverse square law of gravitation. (In this case, the mapping of one ellipse to another 
under $z \rightarrow z^2$ is trivial to understand geometrically.) 

Note that the coupling constants of the two force laws are also naturally mapped as $GM = c_1 E (\ell_G/\ell_H)^2$, where $M$ is the mass of central attracting body in Newton's law of gravity, $E$ is energy of the Hooke's law orbit, $\ell_G$ ($\ell_H$) is angular momentum of gravitational (Hooke's law) problem, and $c_1$ is an arbitrary proportionality constant, inserted so as to take care of dimensions on both sides. Considering the fact that, upon quantization, $E=\hbar \omega (n+1/2)$, it would be interesting to investigate whether the above relation between coupling constants has any deeper significance.

Further discussions on the duality result can be found in \cite{needham-1,needham-2}, and we shall not repeat them here. Rather, we point out how the well-known, conserved Laplace-Runge-Lenz vector for the inverse-square law maps to the Hooke's law case.

\section{The Laplace-Runge-Lenz vector} \label{sec:cons-vec}
Let $Z$ refer to a point on the Keplerian orbit, and $z$ the corresponding point on the Hook's law orbit; therefore, $Z=z^2$. The \LRL vector for the Keplerian case can be written using complex variables as:
\begin{eqnarray}
A = - i {\bar h} P - \mu k \frac{Z}{|Z|}
\label{eq:LRLvec-1}
\end{eqnarray}
where $P=\mu \DM Z / \DM T$, $k=G m_1 m_2$ and $\mu=m_1 m_2 / (m_1+m_2)$, $m_1, m_2$ being the masses of central and orbiting body. Now, let $Z=R \exp{i \theta}$ and $z=r \exp{i \phi}$. Then, the mapping implies, $r=\sqrt{R}, \phi=\theta/2$ [where the choice of phase has the usual ambiguity, and we have chosen the simplest possibility]. From the definition of angular momentum, we have, ${\bar h} \DM T = \mu R^2 \DM \theta$ and $h \DM t = \mu r^2 \DM \phi$. Therefore,
\begin{eqnarray}
A &=& - \mu k \left[ i \left( \frac{{\bar h}^2}{h k} \right) \frac{1}{z^*} \frac{\DM z}{\DM t} + \frac{z}{z^*} \right] \\
&=& - \mu k \left[ i \left( \frac{h}{E} \right) \frac{1}{z^*} \frac{\DM z}{\DM t} + \frac{z}{z^*} \right]
\end{eqnarray}
where we have used the fact that $k=E (\bar h / h)^2$, which follows immediately from Eq.~(\ref{eq:force-law}) with $m=2$. 

This is a very weird looking expression, and it is not straightforward to see that it is indeed a constant of motion for the Hooke's law orbit. In fact, we have not even translated the original \LRL vector appropriately, since, viewed as a complex number, we must have used the inverse map: $A \rightarrow \sqrt{A}$ to map the vector; this is non-trivial, due to the branch cut involved in taking the square root. However, none of these considerations change the fact that the vector represented by the complex number $A$ is constant for a Hooke's law orbit. A direct demonstration of this requires some further work. Letting ${\cal A} = - A / (\mu k)$, we have, using the complex notation (with overdot denoting time derivative),
\begin{eqnarray}
p &=& \mu \dot{z} \;\; ; \;\; \dot p = - \mu \omega^2 z \nonumber \\
p z^* &=& (\bm p \cdot \bm r) + i h
\label{eq:pz-rel}
\end{eqnarray}
Using these, and the fact that $E = |p|^2/(2 \mu) + (1/2) \mu \omega^2 |z|^2$, we can calculate $\dot {\cal A}$ with ${\cal A} = (i h / E \mu) (p/z^*) + (z/z^*)$. We obtain
\begin{eqnarray}
\dot {\cal A} &=& \frac{i h}{E \mu} \underbrace{\left( \frac{\dot p}{z^*} - \frac{p \dot z^*}{{z^*}^2} \right)}_{-2 E / {{z^*}^2} } + \underbrace{\frac{\dot z}{z^*} - \frac{z \dot z^*}{{z^*}^2}}_{\left( p z^* - p^* z \right) / (\mu {z^*}^2)}
\\
&=& 0
\end{eqnarray}
since $p z^* - p^* z=2 i h$ from the last of Eqs.~(\ref{eq:pz-rel}). It is instructive to inspect the relation between the vector $\alpha$ constructed above and the following known conserved quantities for the Hooke's law orbit, as can be found in Sec. 9-7, page 423-424 of \cite{goldst}:
\begin{eqnarray}
S_1 &=& \frac{1}{4 \mu \omega} \left( 2 p_x p_y + 2 \mu^2 \omega^2 x y \right) \\
S_2 &=& \frac{1}{4 \mu \omega} \left( p_y^2 - p_x^2 + \mu^2 \omega^2 (y^2 - x^2) \right) \\
S_3 &=& \frac{h}{2}
\end{eqnarray}
alongwith the condition, $S_1^2 + S_2^2 + S_3^2 = (E/(2 \omega))^2$. Then, by explicit evaluation, it is easy to show that
\begin{eqnarray}
|\mathcal{Q}|^2 &=& \left( \frac{h}{2} \right)^2 \frac{|{\cal A}|^2}{1 - |{\cal A}|^2}
\end{eqnarray}
where $\mathcal Q = -S_2 + i S_1 = \left( p^2 + \mu^2 \omega^2 z^2 \right) / ({4 \mu \omega} )$. It would be interesting to relate the complex number ${\cal A}$ itself (rather than $|{\cal A}|$) to some vector constructed out of $S_1, S_2$.

\section{Concluding Remarks} \label{sec:further}

In standard courses, the subject of classical mechanics is often reduced to mundane equations which are handy in calculating motion of objects under given external conditions. Unfortunately, such a treatment hides some very elegant aspects of the subject, aspects which form a basis for further abstractions and development. As the authors of \cite{sudarshan-mukunda} state in the preface of their book: {\it If true beauty implies that she is ever new, then classical dynamics is truly beautiful.} It has been our endeavour in this note to highlight the geometrical aspects of classical dynamics, by studying the characterisation of a trajectory produced under a given force field in terms of its curvature. We now wish to give a broader overview of such purely geometrical aspects which are related to our discussion here, but whose in-depth study is beyond the scope of this article.

Although we have discussed curves in flat 2-dimensional spaces, more interesting situations arise when these 2-dimensional spaces are themselves embedded in 3-dimensions, which is what one will mostly encounter in physical situations, such as motion of a rigid body in 3-space. (This was in fact the motivation for the so called ``method of moving frames", such as the ``Darboux frame" introduced by geometer Darboux in the late nineteenth century.) In Appendix \ref{app:embedding-of-curves}, we consider this situation, introducing tools from advanced differential geometry in simple and intuitive manner. In particular, Eq.\;(\ref{eq:app-acc-kuu}) is structurally similar to Eq.\;(\ref{eq:force-fn-ft}). We may therefore formalise the notion of curvature of a curve as a measure of bending of the curve. When the curve lies on a surface embedded in 3-dimensional space, there appears a new term (see Appendix \ref{app:embedding-of-curves}), $K_{ab}$, in the relation between intrinsic acceleration and embedding acceleration. However, the 2-surface itself may also have some {\it intrinsic} curvature of it's own, apart from the curvature it inherits from being embedded in a particular way in the 3-space. This intrinsic curvature, which can be measured by measuring angles and distances {\it within} the 2-surface, is denoted by $\cal R$, the so-called Ricci curvature scalar. For 2-dimensional surfaces, there is a beautiful realization due to Carl Friedrich Gauss, his {\it Theorema Egregium} (Latin for ``Remarkable Theorem"), which states that the extrinsic curvature is completely determined by the intrinsic curvature! That is, one can get full information about curvature of curves in a given 2-surface simply by measuring angles and distances within the surface, without having to know how the surface is embedded in 3-space. Specifically, ${\cal R}=2 \kappa_1 \kappa_2$, where $\kappa_1, \kappa_2$ are eigenvalues of the $2 \times 2$ matrix $K_{ab}$ \cite{comment-4}. It is difficult to over-emphasize the importance of this discovery of Gauss, which is often cited as {\it one of the most important and surprising discoveries in all of mathematics} \cite{frankel-theodore}. It explains why one can never make a map of the Earth (a sphere) on a plane sheet of paper without producing distortions, and why this is possible to do for a cylinder \cite{comment-5}.
At a higher level of abstraction, the notion of intrinsic curvature can be introduced for gauge-fields, such as the vector potential in electromagnetism and Yang-Mills theory, in which case this intrinsic curvature measures the strength of the gauge field. We refer the more advanced reader to \cite{frankel-theodore} for a nice introduction to several such topics.

The relevance of geometrical aspects of a trajectory, and the intrinsic curvature of the background on which these trajectories live, is perhaps most particularly evident in the formulation of Einstein's theory of General Relativity (GR), which interprets gravity itself as a manifestation of intrinsic curvature of 4-dimensional space-time, rather than a force as has been looked upon since its introduction by Newton. In a region of space-time small enough so that curvature (measured by the $4^{\rm th}$ rank Riemann tensor) is almost constant, one can make gravity disappear simply by falling freely. Put in this way, the elementary fact that one would feel weightless in a free fall, then leads to an amazing insight in the character of gravitational force. Indeed, this realization of Einstein guided him to the ``principle of equivalence", and at once explained why all masses fall equally fast under gravity -- they do so because all of them are moving in the same geometry, and gravity is geometry. Incidentally, the possibility of getting rid of gravitational effects locally by falling freely suggests that we may regard gravity itself, atleast locally, as a fictitious force, much like the centrifugal and coriolis forces. Indeed, there is deep connection between geometric decomposition of force we have described here and such fictitious forces which arise when working in a non-inertial frame; however, this discussion is out of the scope of present work. Finally, the notion of a ``reference frame" itself, put to good use by Einstein, is based on a one-to-one correspondence between observers and coordinate frames, thereby giving geometrical considerations based on such reference frames an immediate physical perspective. 

Before closing, we cannot resist mentioning that, the two major theories of theoretical physics, GR and Quantum Field Theory, are based, respectively, on geometrization of the $1/r^2$ gravity law (which appears in the weak field limit of GR), and quantization of a collection of harmonic oscillators described by Hooke's law. In the context of the duality between these two highlighted in this note, there may be some further intriguing connections which proper study might unveil. 


\section*{Acknowledgements}
I thank Archana for initial discussions on the topic. I am also grateful to an anonymous referee for several important suggestions which have led to considerable improvement in presentation, in particular: the discussion of force in terms of $F_{_N}$ and $F_{_T}$, inclusion of ``Method 2" to derive Eq.\;(\ref{eq:kappa-rel}), and comments from a broader perspective in the concluding section.

The author's research is funded by National Science \& Engineering Research Council (NSERC) of Canada, and Atlantic Association for Research in the Mathematical Sciences (AARMS).
\appendix
\section{A quick comment on textbook expression for $\kappa$} \label{app:kappa-and-ext-curv}

Let us demonstrate the equivalence of the expression for the Gaussian curvature with the usual one encountered in 
Differential geometry, in the form of Extrinsic curvature associated with embedding of the curve in the Euclidean 
space. Since this stuff is fairly standard, we shall be deliberately brief. It must also be noted that we are 
considering the standard case of a curve in a flat Euclidean space. Since the curve is a one-dimensional manifold, it 
is spanned by one basis vector, which is the tangent vector to that curve. Call this 
$\bm t$. We have, $\bm t \cdot \bm t=1$ and $\bm t \cdot \bm n = 0$. The extrinsic curvature tensor is defined 
as $K_{ij} = \nabla_i n_j - (\bm n \cdot \bm n)^{-1} n_i a_j$, $a^j=n^k \nabla_k n^j$ being the covariant acceleration of $\bm n$. Since 
$K_{ij} n^i = 0 = K_{ij} n^j$, the tensor $K_{ij}$ has components only along the curve. The curve being 
one-dimensional, there is only one component, call it $K$. To find $K$, one projects $K_{ij}$ along the curve: 
$K_{tt}=K=K_{ij} t^i t^j=t^i t^j \nabla_j n_i$, where we have used $\bm t \cdot \bm n = 0$. It is also easy to 
verify that, $t^i t^j = g^{ij} - n^i n^j$ (it is fun to associate this with the more advanced concept of induced 
metric one encounters in standard expositions of Gauss-Codazzi decomposition). Let us indicate why the above relation 
must be true. Choose $\bm t$ and $\bm n$ as basis vectors, so that the corresponding metric components become: 
$g_{tt}=1=g_{nn}, g_{tn}=0$, where $t$ and $n$ are parameters along $\bm t$ and $\bm n$. 
An arbitrary displacement in arbitrary coordinates can be written as $\bm \DM x = \bm t \DM t + \bm n \DM n$. 
Therefore the line element becomes: $\DM s^2 = \DM t^2 + \DM n^2 = (\bm t \cdot \bm \DM x)^2 
+ (\bm n \cdot \bm \DM x)^2 = (t^i t^j + n^i n^j) \DM x^i \DM x^j$, which is the desired result. 

Thus, we finally obtain, $K=g^{ij} \nabla_j n_i=\nabla_i n^i$. The equivalence with standard result in analytic 
geometry is then established by realizing that, $K=t^i t^j \nabla_j n_i$ is essentially 
$\bm t \cdot \left( {\DM \bm n}/{\DM s} \right) = - \bm n \cdot \left( {\DM \bm t}/{\DM s} \right) = \kappa$, where $s$ is the 
arclength parameter along the curve.

\section{Curvature of Curves and Embedding} \label{app:embedding-of-curves}

We clarify the dependence of the Gaussian curvature on the intrinsic curvature of the surface on which the curve lives, for example, a curve on a sphere. (In what follows, we shall drop the hat on the unit vector $\bm n$ for notational clarity, and write $n^2=1$ explicitly so as to facilitate a direct generalization to the case where $n^2=\pm 1$, such as in Lorentzian spacetimes.) Intuitively, we can easily see this dependence for a sphere as follows. The equator, which is a geodesic on the sphere, is the straightest possible curve on the sphere, but viewed as a circle in flat, Euclidean space, it is a curve of constant Gaussian curvature. We will establish here several relations that quantify this intuitive picture. The only concept from advanced differential geometry that is needed here are the Gauss-Weingarten equations \cite{mtw-Kab2}, the form of which is intuitively easy to understand.
\begin{eqnarray}
\bm \nabla_{\bm e_{(b)}} \bm e_{(a)} = \Gamma^c_{a b} \bm e_{(c)} - K_{a b} \frac{\bm n}{n^2}
\label{eq:gauss-weingarten}
\end{eqnarray}
with following meanings: $\bm \nabla$ stands for covariant derivative with respect to a $D$-dimensional manifold 
$\mathcal M$, into which is embedded a $(D-1)$-dimensional curved manifold $\mathcal S$ [such as a sphere in a $3$-$D$ 
Euclidean space]; the indices $a, b \ldots$ refer to $\mathcal S$, and $\Gamma^c_{a b}$ is the Christoffel connection 
appropriate to $\mathcal S$. The embedding is defined by giving the set of orthogonal vectors, 
$\{ \bm e_{(a)}, \bm n \}$, where $\bm e_{(a)}$'s span $\mathcal S$ and $\bm n$ is normal to $\mathcal S$. Now 
consider a curve in $\mathcal S$, with tangent vector $\bm u = u^a \bm e_{(a)}$. Its acceleration as seen from 
$\mathcal M$ is $\bm A_{\mathcal M} = \bm \nabla_{\bm u} \bm u$, while as seen from $\mathcal S$, it will involve the 
covariant derivative defined with respect to $\Gamma^c_{a b}$. We obtain
\begin{eqnarray}
\bm A_{\mathcal M} &=& \bm \nabla_{\bm u} \bm u 
\nonumber \\
&=& u^b \bm \nabla_{\bm e_{(b)}} \bm \l( u^a \bm e_{(a)} \r)
\nonumber \\
&=& u^a u^b \bm \nabla_{\bm e_{(b)}} \bm e_{(a)} + \bm e_{(a)} u^b e_{(b)}^j \partial_j u^a
\nonumber \\
&=& u^a u^b \bm \nabla_{\bm e_{(b)}} \bm e_{(a)} + \bm e_{(a)} u^j \partial_j u^a
\nonumber \\
&=& u^a u^b \l\{ \Gamma^c_{a b} \bm e_{(c)} - K_{a b} \frac{\bm n}{n^2} \r\} + \bm e_{(a)} u^j \partial_j u^a
\nonumber \\
&=& \bm A_{\mathcal S} - K_{uu} \frac{\bm n}{n^2} \label{eq:app-acc-kuu}
\end{eqnarray}
where $K_{uu}=K_{a b} u^a u^b$. Let us explain the relevant steps a bit. The 2nd equality follows from the linearity 
of covariant derivative operator. In the 3rd equality, the partial derivative occurs because $u^a$'s are scalars in 
the embedding space, with respect to $\bm \nabla$. To arrive at the final equality from the previous line, we have 
used the fact that, $u^j \partial_j u^a = \DM u^a / \DM \tau$ (chain rule), which combines with the first term and is 
seen to be the definition of acceleration of the curve as defined using the connection $\Gamma^c_{ab}$ appropriate 
to $\mathcal S$.

Taking the magnitude, we obtain
\begin{eqnarray}
A_{\mathcal M}^2 = A_{\mathcal S}^2 + \frac{1}{n^2} K_{uu}^2
\label{eq:acc-embedding}
\end{eqnarray}
This is an extremely interesting relation, since it relates the acceleration [and hence curvature] of the same curve 
as embedded in two different manifolds, in a Pythagorean manner. Incidentally, there is also an immediate 
generalization to the case when $\mathcal S$ is $(D-m)$-dimensional, where $m<D$. In this case, the right hand side 
above will involve the sum of squares of $``K u u"$ for all $\bm n_{(k)}$, $k=1 \ldots m$.

We can in fact simplify the $K_{uu}$ term further. From Eq.~(\ref{eq:gauss-weingarten}), it is easy to see that 
(the capitalised indices $I,J$ represent components w.r.t. ${\mathcal M}$; for e.g., $u^J=u^a e_{(a)}^J$)
\begin{eqnarray}
K_{ab} &=& -n^I e_{(b)}^J \nabla_J e_{(a) I}
\nonumber \\
K_{uu} &=& -n^I u^a u^J \nabla_J e_{(a) I}
\nonumber \\
&=& u_I u^J \nabla_J n^I
\nonumber \\
&=& - n^I u^J \nabla_J u_I
\nonumber \\
&=& - \bm n \cdot \bm A_{\mathcal M} 
\end{eqnarray}
Using this, Eq.~(\ref{eq:acc-embedding}) can be rewritten as
\begin{eqnarray}
A_{\mathcal S}^2 = A_{\mathcal M}^2 \l[ 1 - \l( \bm{{\hat A}}_{\mathcal M} \cdot \bm {n} \r)^2 \r]
\end{eqnarray}
We can now use this relation to confirm our intuition in the case of equator on a sphere. This can, in fact, be done 
purely geometrically. In the standard spherical coordinates, for any latitude other than the equator, the acceleration 
vector as seen from flat space will be directed outwards perpendicular to the $Z$-axis and in the $X-Y$-plane, whereas 
$\bm n$ is obviously $\bm {\hat r}$. The angle between them is $(\pi/2) - \theta$, so that we obtain 
$A_{\mathcal S}^2 = A_{\mathcal M}^2 \cos^2 \theta$, which vanishes, as expected, for the equator $\theta=\pi/2$. 
A little imagination is enough to convince one that $A_{\mathcal S}$ would vanish for any great circle on the sphere, as befits a geodesic. 

Finally, let us quickly confirm the above relation using brute force techniques of elementary differential geometry. A 
sphere is a $r=$constant surface in $3$-$D$ space. The metric is, $\DM s^2 = \DM r^2 + r^2 \l( \DM \theta^2 
+ \sin(\theta)^2 \DM \phi^2 \r)$. Consider now any latitude on the sphere, defined by $\theta=$constant in addition to 
$r=$constant. The unit tangent vector (in $(r,\theta,\phi)$ coordinates) is $u^k=[0,0,1/(r \sin \theta)]$. Its 
acceleration is $A_{\mathcal M}^k=u^j \nabla_j u^k=-(1/r)[1,\cos \theta / (r \sin \theta),0]$. Note than, upon 
transforming to Cartesian coordinates we have $A_{\mathcal M}^Z=0$, which confirms what was said in the earlier 
paragraph. Therefore,  $A_{\mathcal M}^2 = 1/(r^2 \sin^2 \theta)$. (Note that $r \sin \theta$ is the radius of the 
circle.) On the other hand, the metric on the sphere itself is $\DM s^2 = r^2 \l( \DM \theta^2 + 
\sin(\theta)^2 \DM \phi^2 \r)$, and $u^k=[0,1/(r \sin \theta)]$ in $(\theta, \phi)$ coordinates. We obtain, 
$A_{\mathcal S}^k=u^j D_j u^k=-(1/r)[\cos \theta / (r \sin \theta),0]$, where $D$ is covariant derivative w.r.t metric on the sphere. Therefore, $A_{\mathcal S}^2 = \cos^2 \theta/(r^2 \sin^2 \theta)$. Hence, we immediately see 
that $A_{\mathcal S}^2 = A_{\mathcal M}^2 \cos^2 \theta$, as was derived in the previous paragraph.



\begin{thebibliography}{99}
\bibitem{comment-1}
Note that in the standard definition, there is no minus sign. Our convention for the normal vector, shown in Fig.~\ref{fig:set1}, is opposite to the standard convention, in which case the normal is inward pointing.
\bibitem{comment-2}
There is a very crucial point we must keep in mind: all our analysis here is in two dimensions, and is applicable only to those three-dimensional motions that are confined to a plane (which is the case for the force laws we will consider due to conservation of angular momentum).
\bibitem{comment-3}
We must parallel transport the vectors to the same point to compare them, but because we are in flat Euclidean space, the transport is trivial.
\bibitem{comment-4}
In matrix notation, the relation ${\cal R}=2 \kappa_1 \kappa_2$ can also be written as ${\cal R}=({\rm Tr}\,K)^2 - {\rm Tr}\,K^2$, where ${\rm Tr}$ is the trace operation.
\bibitem{comment-5}
The explanation goes as follows: a map without distortion must preserve distances and angles, or $\cal R$ (such a map is called isometric) -- however, ${\cal R}=0$ for a plane and a cylinder, while it is non-zero for a sphere, for which $\kappa_1=1/a=\kappa_2$ ($a$ is the radius of the sphere). Hence the inevitable distortion in any planar map of the earth, something the ancient cartographers were well aware of!
\end{thebibliography}
\end{document}